\documentclass{Rinton-P9x6}
\usepackage{latexsym,amssymb,amsmath}

\begin{document}

\title{Will LIGO see RIGO's? --- Radion-induced graviton oscillations in
the two-brane world}

\author{Andrei O.\ Barvinsky}
\address{Theory Department, Lebedev Physics Institute,
Leninsky Pr.\ 53, Moscow 117924, Russia
\\E-mail: barvin@td.lpi.ru}  

\author{Alexander Yu.\ Kamenshchik}
\address{L.\ D.\ Landau Institute for Theoretical Physics
of Russian 
Academy of Sciences, Kosygina str.\ 2, Moscow 117334, Russia;\\
Landau Network --- Centro Volta, Villa Olmo, via Cantoni 1, 22100 Como, 
Italy
\\E-mail: sasha.kamenshchik@centrovolta.it}

\author{Claus Kiefer}
\address{Institut f\"ur Theoretische Physik, Universit\"at zu K\"oln,  
Z\"ulpicher Str.\ 77, 50937~K\"oln, Germany
\\E-mail: kiefer@thp.uni-koeln.de}

\author{Andreas Rathke}
\address{ESA Advanced Concepts Team (SER-A), ESTEC, Keplerlaan 1,
2201 AZ Noordwijk, The Netherlands
\\E-mail: andreas.rathke@esa.int}


\maketitle

\abstracts{ One of the most interesting features of braneworld models is the
  existence of massive gravitational modes in addition to the usual massless
  one. Mixing of the modes which depends nontrivially on the inter-brane
  distance can be interpreted as radion-induced gravitational-wave
  oscillations, a classical analogy to meson and neutrino oscillations. We
  show that these oscillations arising in M-theory inspired two-brane setups
  could lead to effects detectable by gravitational-wave
  inter\-fe\-ro\-meters.}


Massive gravitational modes are a generic feature of higher-dimensional
spacetime models.  They arise from the expansion of the linearized
gravitational modes of the full spacetime, the bulk, in harmonics of the
higher dimensions.  By the expansion one obtains one mode with zero
eigenvalue, the massless four-dimensional graviton,\footnote{Adopting to
common jargon we will denote by graviton the classical linearized modes of
the gravitational field.} 
and an infinite set of modes with non-zero
eigenvalues, the massive Kaluza-Klein (KK) tower of gravitons.  In models
based on compactifications of flat extra dimensions on a circle the KK
modes are of little relevance for the low-energy four-dimensional dynamics
because the modes are too
heavy to be produced by astrophysical sources. 
In these models the coupling of all modes to matter is of the same strength.

The situation is different in scenarios with warped extra dimensions, i.\,e.
scenarios in which the extra dimension has constant curvature induced by a
bulk cosmological constant and in which the matter of our Universe is usually
confined to a brane. Due to the bulk curvature the four-dimensional effective
coupling of gravitation to matter on the brane depends on the position of the
brane in the bulk. In particular, the eigenmodes of the higher-dimensional
harmonics will in general couple to matter on the brane with unequal strength.
This property considerably relieves the constraint on the size of the extra
dimension coming from the lack of evidence for the generation of KK modes and
thus allows the KK modes to be much lighter than in scenarios with flat higher
dimensions. Hence, in warped compactifications part of the gravitational
interaction between matter on the brane can be mediated by the KK modes,
although their contribution to the gravitational force will be small because
their coupling to matter is suppressed compared to the massless mode.

Accordingly, also high-frequency astrophysical sources of
gravitational waves (GW's) will release a small fraction of their radiation as
massive GW's. 
The co-occurrence of the massive GW's with the massless GW at first sight seems
to complicate the detection of the KK modes but actually turns out useful for
observational purposes:\footnote{We do not discuss here their different
  polarization properties associated with the different number of
  polarizations of massless and massive modes.} due to the different
propagation speeds of waves with different masses the superposition of the
waves leads to a beat in the intensity of the total GW.  Its oscillation
length and amplitude of the beat depend parametrically on the size of the
extra dimension, which is described by the radion field,
thus suggesting the name radion-induced graviton
oscillations\cite{rigoshort,rigolong} (RIGO's).

RIGO's can in principle occur in any model with gravitons of different masses
although only in models with warped compactifications they are likely to occur
in GW's from astrophysical sources.  For simplicity, we discuss RIGO's in the
Randall-Sundrum (RS) two-brane model\cite{RS1} although other --- more
conveniently tuned --- models might be more favorable from the observational
point of view. The RS two-brane model is a static solution of the
five-dimensional Einstein equations with an anti-de\,Sitter (AdS) bulk which
is compactified on a $S^1/{\mathbb Z}_2$ orbifold. Each brane is locate on one
of the orbifold fixed planes. The brane tensions $\sigma_\pm$, being of equal
magnitude but of opposite sign, and the bulk cosmological constant are
fine-tuned to yield conformally-flat induced metrics on the branes (we chose
the metric on the positive tension brane to be the Minkowski metric). The
conformal factor relating the induced metrics on the branes $a$, usually
called the warp factor, can be chosen arbitrarily. Considering it as a
dynamical variable, it can be identified as the field determining the brane
distance, the radion. In the following we take the point of view that the
positive-tension ($\sigma_+$-) brane will be the phenomenologically- as well
as M-theoretically-favored brane to be identified with our
Universe.\cite{Pyr,rigolong}

The linearized effective four-dimensional equations of motion for
transverse-traceless perturbations, i.\,e.\ GW's, on the
$\sigma_+$-brane are conveniently written in their spectral representation in
terms of four-dimensional mass-eigenmodes. This representation can be obtained
e.\,g.\ 
by localizing the underlying
effective nonlocal action\cite{nlbwa,rigoshort,rigolong}. Suppressing tensor
indices, we have for the metric perturbation $h^+$ on the $\sigma_+$-brane
\begin{equation}
       h^+= - 16\pi G_4 \Bigg[
\frac{1}{\Box} 
(T^++a^2 T^-) \\ 
        + \sum^\infty_{i=1}
\frac{a^2}{\Box-m_i^2} 
\left( \frac {T^+ }{\mathcal{J}_{2,i}^2}- 
       \frac {T^-}{\mathcal{J}_{2,i}}\right) \Bigg] \, ,
\label{eom}
\end{equation} 
where $1/\Box$, $1/(\Box-m_i^2)$ denote massless and massive retarded scalar
Green functions, respectively. In Eq.\ (\ref{eom}) we use the abbreviation
$J_{2,i} \equiv J_2(lm_i/a)$ for the value of the Bessel function of second
order with the argument $lm_i/a$, $l$ being the AdS radius of the bulk and
$m_i$ the mass of the $i$th KK mode. We consider an idealized astrophysical
source at ${\mathbf x}=0$ on each brane with a harmonic time dependence,
$T^\pm(t,{\mathbf x})=\mu e^{-i\omega t}\delta({\mathbf x})$, where $\mu$ is
the quadrupole moment of the source. For further simplification we consider a
source frequency $\omega$ above the mass threshold of the first massive mode
but below the threshold of the second KK mode, i.\,e.\ $m_1 < \omega < m_2$.
Then only the massless and the first massive mode are excited and produce
long-range GW's.  At a distance $r$ from the source the waves on the
$\sigma_+$-brane are given by a mixture of massless and massive spherical
waves,
 \begin{flalign} 
 h^+[\,T^+] & = A\, e^{-i\omega t} 
 \left(e^{i \omega r} +  a^2/\mathcal{J}_2^2  
 e^{i \sqrt{ \omega ^2-m_1^2}r} 
                              \right) \, ,
\label{h+T+} \\ \label{h+T-}  
 h^+[\,T^-] & = A a^2\, e^{-i\omega t}  
 \left( e^{i\omega r} - 1/\mathcal{J}_2  
 e^{i \sqrt{\omega ^2-m_1^2}r} \right) \, ,          
 \end{flalign} 
 for the sources $T^\pm$ on the $\sigma_+$- or $\sigma_-$-brane, respectively,
 where $A=4\,G_4\,\mu/r$ is the amplitude of the massless mode generated by
 $T^+$. In Eqs.\ (\ref{h+T+}) and (\ref{h+T-}) we have introduced
 $\mathcal{J}_2 \equiv \mathcal{J}_{2,1} \approx 0.403$.



From the structure of Eqs.\ (\ref{h+T+}) and (\ref{h+T-}) one realizes that
the waves from both sources exhibit a beat with the oscillation length
\begin{equation} 
L= 2\pi / \Big(\omega - \sqrt{\omega ^2-m_1^2}\Big) \, .\label{length} 
\end{equation} 
The maximal amplitudes $\mathcal{A}^\pm$ of the waves (\ref{h+T+}) and
(\ref{h+T-}) are given by
\begin{equation} 
\mathcal{A}^+ = \left(1+ a^2 / \mathcal{J}^2_2 \right) A \approx A \, , 
\quad \quad 
\mathcal{A}^- = \left(1 / \mathcal{J}_2 -1\right) a^2 \,A 
\approx  1.5 \, a^2A \, , 
\label{prefacA} 
\end{equation} 
where the approximations are valid in the limit $a\ll 1$. For a GW produced by
$T^+$, Eq.~(\ref{h+T+}), the amplitude modulation of the beat is suppressed by
a factor of $a^2$ compared to $\mathcal{A}^+$ in the limit of large brane
separation, $a \ll 1$. The amplitude of the GW produced by $T^-$, Eq.\ 
(\ref{h+T-}), is dominated by oscillation of the beat, regardless of the
inter-brane distance.

In the limit of a source-frequency just above the mass-threshold 
$\omega \gtrsim m_1$ the oscillation length (\ref{length}) of these
gravitational (``graviton'') wave oscillations tends to
\begin{equation}
L \gtrsim 2\pi / m_1 \approx 1.6\, l / a \, .
\label{short-length}
\end{equation}
The minimal oscillation length (\ref{short-length}) is inverse proportional to
$a$.  Hence, both characteristic quantities of the oscillation, the amplitude
and the oscillation length depend nontrivially on the radion, justifying the
designation RIGO.  RIGO's become observable when their oscillation length is
at least of the order of the arm length of a GW detector. For the ground-based
interferometric detectors this requirement corresponds to $L \sim 10^3\,{\rm
  m}$.  Combining this with the constraint on the maximal size of the AdS
radius $l$ from sub-millimeter tests of gravity, $l \lesssim 10^{-4}\,{\rm
  m}$,\cite{Gundlach} we find an upper limit on the warp factor $a \lesssim
10^{-7}$ for the oscillation length to be
detectable.  Hence, for the ratio of the amplitudes 
of Eq.\ 
(\ref{prefacA}), we find
\vspace*{-2mm}
\begin{equation} 
\mathcal{A}^- / \mathcal{A}^+  \sim a^2 \lesssim 10^{-14}. 
\label{damping}
\end{equation} 
Therefore, the amplitude of a wave originating from a source on the
(``hidden'') $\sigma_-$-brane with oscillations which are sufficiently long to
be detectable, is strongly suppressed by a damping factor $a^2$ compared to a
GW generated by a source on the $\sigma_+$-brane itself.  A strongly
oscillating wave has to be generated by a source 14 orders of magnitude
stronger than that of a weakly oscillating one in order to be of the same
magnitude, which at first sight makes the detection of RIGO's in the RS model
impossible.

The strong suppression of RIGO's found above does not yet take into account
any specific production mechanism of GW's. If one considers the RS two-brane
model as a toy-model for an M-theory realization of large extra dimensions one
finds a generation mechanism for GW's on the $\sigma_-$-brane which naturally
compensates the damping factor found above.\cite{rigolong} In cosmological
M-theory models, the Calabi-Yau (CY) volume $V$ is smaller at the position of
the hidden brane than at the visible brane.\cite{Witten} For example, in the
M-theory solution of \cite{LOSW} the ratio of the CY volume at the two branes
depends on the warp factor as $V^-/V^+ \sim a^6$.  The vacuum expectation
value (VEV) of the gaugino condensate on the hidden brane, $\eta$, depends on
$V^-$ as $\eta \sim {(V^-)}^{-9/2} \exp (- V^- \, \mathcal{S})$, where
$\mathcal{S}$ is a positive function of the CY moduli, the unified gauge
coupling and its renormalization-group $\beta$-coefficient.\cite{gaugino}
Therefore, $\eta$ may become very large when $V^-$ becomes small.  On the
other hand, the amplitude of GW's radiated by cosmic strings is proportional
to the VEV squared of the associated symmetry breaking because the quadrupole
moment of a cosmic string is given by $\mu = \Gamma \eta^2 / \omega^{3}$,
where $\Gamma \approx 50 \ldots 100$ is a numerical coefficient depending on
the trajectory and shape of the string loop, and $\omega$ is the
characteristic frequency of string oscillations.\cite{VilShel} Thus the large
gaugino VEV can lead to the production of strong-amplitude GW's on the hidden
brane, easily compensating the damping factor (\ref{damping}).

Above we found that the different speeds of propagation of the zero mode and
the KK modes of GW's in the RS model induce a beat in the amplitude of GW's.
It is obvious that such RIGO's are completely generic and are a feature of
any higher dimensional model with compactified extra dimensions.

The obvious question to ask is: what would be the concrete signature of RIGO's
in a GW interferometer? The pessimistic answer is that RIGO's will probably
only show up in the stochastic GW background. However, even with the detector
templates for a network of cosmic strings at hand it is still tempting to
detect RIGO's because we have next to no theoretical input for estimating the
warp factor $a$ which governs both the amplitude and oscillation length of
RIGO's.  Thus, although possibly within reach of the first or second
generation of GW interferometers, the detection of RIGO's may only become
possible after we have stronger theoretical constraints on the allowed
parameter space of large extra dimensions. The optimistic answer is that,
given a reliable theoretical prediction for the warp factor, one could easily
extend existing detector patterns to incorporate the effect of RIGO's. The
characteristic oscillation pattern of RIGO's could then even become an
important effect to differentiate between the stochastic GW background and
detector noise and to demonstrate the existence of extra dimensions.

\section*{Acknowledgments}

One of the authors (A.R.) has benefitted from helpful discussions with G.\ J.\ 
Mathews. The work of A.O.B.\ was supported by the Russian Foundation for Basic
Research under the grant No 02-01-00930 and by the grant of Leading Scientific
Schools 1578.2003.2, while A.Yu.K.\ was supported by the RFBR grant No
02-02-16817 and by the scientific school grant No 2338.2003.2 of the Russian
Ministry of Industry, Science and Technology.  A.R.\ was supported by the DFG
Graduiertenkolleg ``Nonlinear Differential Equations'' Freiburg.

\end{document}